Original Paper

# BIG-AOME: Designing Bodily Interaction Gamification towards Anti-sedentary Online Meeting Environments

Jiaqi Jiang[1†], Shanghao Li[1†], Xian Li[1], Yingxin Xu[1], Jian Zhao[2], Pengcheng An[1*]

[1]Southern University of Science and Technology

[2]University of Waterloo

## Abstract

**Background:** Online meetings have become an integral part of daily life for many people. However, prolonged periods of sitting still in front of screens can lead to significant, long-term health risks. While previous studies have explored various interventions to address sedentary lifestyles, few have specifically focused on mitigating sedentary behavior during online meetings. Furthermore, design opportunities to address this issue in the context of online meetings remain underexplored.

**Objective:** This study aims to investigate the design of effective anti-sedentary interactions for online meeting scenarios and understand user experiences with gamified bodily interactions as an anti-sedentary measure during online meetings.

**Methods:** This study adopts a "research through design" approach to develop and explore user experiences of gamified bodily interactions as interventions to mitigate sedentary behavior during online meetings. In collaboration with 11 users, we co-designed and iterated three prototypes, which led to the development of BIG-AOME (Bodily Interaction Gamification towards Anti-sedentary Online Meeting Environments) framework. User studies were conducted with three groups totaling 15 participants, utilizing these prototypes. During co-design and evaluation, all group semi-structured interviews were transcribed into written format and analyzed using Hsieh's conventional qualitative content analysis method.

**Results:** The findings demonstrate that gamified bodily interactions encourage users to engage in physical movement while reducing the awkwardness of doing so during online meetings. Seamless integration with meeting software and the inclusion of long-term reward mechanisms can further contribute to sustained usage. Additionally, such games can serve as online icebreaker tools or playful tools for decision-making. Drawing from three design prototypes, this study offers a comprehensive analysis of each design dimension within the BIG-AOME framework: body engagement, attention, bodily interplay, timeliness, and virtual and physical environments.

**Conclusions:** Our research findings indicate that designing anti-sedentary bodily interactions for online meetings has the potential to mitigate sedentary behaviors while enhancing social connections. Furthermore, the BIG-AOME framework that we propose explores the design space for anti-sedentary physical interactions in the context of online meetings. This framework detailing pertinent design choices and considerations.

# Introduction

## Background

Online meetings have become a significant aspect of modern work, introducing new challenges into daily health. An increasing number of activities are being conducted primarily or entirely online [1]. As noted by Wu and Yu [2], individuals tend to prefer online meeting platforms due to advantages such as flexibility and cost-effectiveness, even when working from home is not mandatory. However, increasing research evidence has revealed various kinds of technostress in virtual conferencing, which can negatively impact mental and physical well-being, productivity, job satisfaction, and group commitment [1,3–5]. Among these challenges, prolonged sedentary behavior stands out as a significant threat, potentially causing irreversible health damage over time. [6–8].

Sedentary behavior refers to periods when the body remains awake but motionless, with activities such as sitting, lying, or reclining[9]. Distinctive from merely a lack of physical activity, the negative effects of prolonged sedentary periods cannot be fully mitigated by intermittent physical exercise [10]. According to current medical studies, there is no established "gold standard" for measuring sedentary behaviors nor specific guidelines on how to effectively interrupt such behaviors [11]. The prevailing recommendation is to reduce excessive sedentary behavior and substitute it with physical activities ranging from mild to intense in nature [12,13]. Seated activities that incorporate moderate movements of the lower or upper body can help reduce the negative effects of sedentary behavior [12]. This approach underscores the necessity to integrate more active routines in daily life to counteract the health risks associated with prolonged sedentary periods.

Many interventions have been developed to lessen sedentary time and encourage more active engagement in physical activities [14,15]. These include break reminder software for computers [16,17], mobile applications designed to interrupt prolonged sedentary behavior[18], dedicated devices equipped with LED displays [19], posture-adjusting smart chair [20], and indoor location-based mobile games [21]. However, online meeting environments, as distinct sources of technostress due to remote social interactions, remain underexplored in design research. Therefore, there is a pressing need for research to collect a broader range of design examples and provide specific guidance for developing effective and engaging interventions to address sedentary behavior in online meetings.

In-person meetings typically feature habitual breaks, encouraged by social and environmental elements [22] . However, online meetings often lack these cues, or social expectations, leading to less affordances for breaking sedentary routines [23]. This indicates the need for designing anti-sedentary activities that could fit into online meetings and engage users in a socially amusing way. Analysis of large-scale video-meeting data shows that multitasking is a common behavior during online meetings. While multitasking can lead to negative outcomes, such as increased distraction [24], it may also have positive effects, such as improved efficiency in attention division or task shifting [25]. This highlights a design opportunity to harness the positive aspects of multitasking in online meetings to reduce sedentary behavior by incorporating physical exercises into the meeting process as secondary or parallel activities. Gamified bodily interactions offer a promising approach to integrating simple, brief physical exercises into session routines.

### Gamification and Exertion Games

Gamification is defined as "the use of game elements in non-game contexts" [26]. It applies principles of game design and elements such as storytelling, leaderboards, and winning rules to address real-world challenges in areas like training, healthcare, and education [27]. Previous studies have demonstrated that gamification can facilitate behavior change[28–30] and make products more engaging [31]. Exertion games/Exergames [32], which promote physical activity by incorporating exercise into digital games, have become an increasingly relevant area of research. According to Muller et al [33], all computer systems that facilitate physical exertion as part of the interaction could be regarded as exertion games, no matter if they are more games focused or exercised-focused.

Many studies have focused on motivating physical activity through gamification by developing software or using hardware [34–39]. Exertion games have demonstrated not only their effectiveness in promoting physical activity but also their ability to enhance social interaction among users [40,41]. Mandryk et al [42] analyzed exertion games from the lens of mitigation of sedentary behaviors, and articulated two design principles: *"Providing an easy entry into play"* and *" Motivating repeated play sessions throughout the day"*. Gamified bodily interactions have been applied in areas such as fitness, education, and training [30,43,44]. However, to the best of our knowledge, few studies have investigated how gamified bodily interactions can be integrated into virtual conferencing scenarios to effectively promote physical activity and combat sedentary behavior.

At the intersection of remote education and exertion games, Shin et al [45] proposed Jumple, a virtual physical education classroom that leveraged AI pose estimation technology to facilitate physical activity for students in online environments. By leveraging commonly available devices, Jumple addresses the challenges of remote learning and the negative impact of the pandemic on children's well-being. Sachan et al [46] investigated the use of AR-based micro health interventions to mitigate Zoom fatigue among college students during virtual classes, exploring their effectiveness in reducing sedentary behavior and promoting physical activity. These studies offer design instances from online classroom settings, which is a similar context to online meetings. Building further upon them, our exploration pertinently focuses on the online meeting context and aims to further generate knowledge in the form of a design framework, to systematically surface the design space and implications.

## Understanding the Online Meeting Environment

For decades, scholars have predicted that online meeting will reform the conventional routine of commuting to and from workplace and the way people collaborate with others [47]. Yet, for many people, the heightened reliance on online meetings has proven to be mentally and physically challenging. This is described as "Zoom fatigue", an emerging term and phenomenon to denote the common fatigue associated with videoconferencing [5]. Rudnicka et al [48] found that break-taking is a behavior mediated by the social norms among coworkers, while remote work environments often lead to extended periods of inactivity and excessive work due to the absence of social cues. Given the drawbacks and challenges associated with current videoconferencing software, we propose to add socially engaging bodily interactions to afford more anti-sedentary components in online meetings. Many people perceive unhelpful breaks as those involving sedentary screen activities (e.g., games, social media, web browsing, video watching), and helpful breaks being physical activities [49]. However, in practice, people tend to exceed the intended duration for digital and static breaks, whereas physical breaks, especially outdoor activities are less likely to extend beyond the planned time [40]. Facing this situation, we explore the potential of adding low-threshold physical movements as a playful anti-sedentary option. Serving as a supplement for higher-threshold physical activities (e.g., leaving the seat or walking around), these games offer an additional option promoting helpful breaks during online meetings.

Multitasking is a common phenomenon during meetings. In face-to-face meetings, people engage in various physical activities such as pacing, standing, and stretching, which are beneficial to overall meeting performance[51]. During phone calls, people also tend to perform minor movements while talking without disrupting the communication [47]. However, online meetings include more digital multitasking than physical multi-tasking (e.g., locomotion or other movements). On a Zoom call, maintaining a central position within the camera view with one's face clearly visible to others is regarded by cultural norms as professional and trustworthy [52]. Research often stresses the negative impacts of multitasking including increased mental workload and reduced productivity, which could potentially contribute to burnout and depression [5]. However, according to a large-scale analysis of remote meeting multitasking behaviors, in-meeting multitasking can also lead to positive outcomes when participants are able to regulate their attention, making flexible use of time when a meeting part is not demanding or critical to them [25]. The authors suggest we should allow *"space for positive multitasking"* and *"shorten meeting duration and insert breaks"*. In the future, people's tendency for multitasking is unlikely to diminish due to the growing prevalence of digital devices and the nature of the modern workplace [53]. Inspired by the idea of leveraging positive multitasking to mitigate sedentary routines, our design incorporates bodily game elements into the online meeting interface.

## Objectives

Given that little knowledge has been accumulated regarding *how to design gamified bodily interactions as anti-sedentary interventions in online meetings*, our study has

adopted a Research-through-Design (RtD) methodology [54] to address two research questions:
- Research Question 1 (RQ1): how can gamified bodily interactions be integrated into online meetings to reduce sedentary behavior?
- Research Question 2 (RQ2): what are the relevant design options and considerations to properly design such gamified bodily interactions for online meeting contexts?

To address the first part of this research, we conducted three rounds of co-design activities with 11 participants. During these sessions, participants brainstormed, elaborated, and assessed initial bodily game ideas and demos considering their prior online meeting experiences. Based on the co-design insights and outcomes, we propose the initial BIG-AOME framework that describes a design space to help designers/researchers consciously navigate multiple relevant design dimensions. Across these dimensions, the BIG-AOME framework reveals various design options that can be considered to make thoughtful design choices according to specific design goals and scenarios.

To further consolidate and contextualize the BIG-AOME framework, in the second part of this research, we implemented three game design ideas into functioning prototypes and evaluated them with 15 participants in online meeting settings. The three designs were chosen because they were hypothesized to represent distinct design options across various dimensions, enabling a concrete exploration of the design space and uncovering context-specific insights and implications underlying the BIG-AOME framework. These bodily game prototypes can be integrated into mainstream online meeting platforms (e.g., Zoom or Voov Meeting) via virtual camera software OBS [55] . The visual components of the games are overlaid on the camera view or video tiles of online meeting attendees. While the games are launched, the attendees could interact with each other and the game objects via bodily movements adapted from existing beneficial physical exercises. Thus, gamified exercises could be embedded flexibly into meeting sessions as a playful way to mitigate sedentary meeting routines, with the participants still being seamlessly engaged in the meeting and connected with other attendees.

Our study offers a design-oriented exploration of gamified bodily interactions as a novel avenue to break sedentary online meeting routines. Our contributions are twofold: (1) three prototypes as design instances of anti-sedentary bodily gamification for online meetings, along with empirical findings into user experiences; and (2) a preliminary design framework for creating gamified bodily interactions for online meetings, surfacing a promising design space with relevant design options and considerations to inform and inspire future research.

## Methods

### Design Exploration and Formulation of the BIG-AOME Framework

In this section, we present our iterative research-through-design approach [54], including our early design exploration, co-design workshops, and the initial formulation of the BIG-AOME design framework.

#### *Early Design Explorations*

In the initial stage, we explored the possibility of integrating gamified bodily interactions as anti-sedentary interventions in online meeting scenarios. To do so, we first studied existing literatures [26–30] and online resources (e.g., credible health information websites such as MedlinePlus[a], Centers for Disease Control and Prevention[b], and World Health Organization[c] and YouTube channels maintained by professional physical therapists or exercise coaches). The purpose was to accumulate suitable anti-sedentary physical exercises as design inputs. These exercises specifically target areas such as the shoulders, neck, and back, all of which can be performed in typical online meeting settings. This corpus of verified healthy movements served as an inspiration resource for gamification design ideas to build upon. Following this, we generated a wide range of gamification ideas, implemented them into interactive demos, and tested these for firsthand experience and as preparatory materials for later co-design workshops.

*Insights from early design exploration*: The most important lesson we learned was that simply using bodily interactions to control a "classic" digital game would often fail to provide appropriate gaming experiences. Two failed examples from our early attempts were arm-controlled Angry Birds and Snake. Both games were considered blunt and tedious when using arm gestures to control and they caused body fatigue instead of mitigating the tension from sedentary behaviors. According to Mueller et al.[56], these designs used the body as merely a controller for digital objects, thus failing to fulfill the advantages of "body as play". Instead, we focused on creating gameplay experiences that could naturally embody the players in the virtual meeting environments to achieve in-depth and seamless bodily engagement.

The early design exploration laid the groundwork for our BIG-AOME framework. In the initial stages, we developed design considerations based on literature research, online resources, and insights from early prototyping. These considerations addressed various aspects and formed the initial elements of our framework. In the subsequent co-design sessions, we refined these considerations based on users' needs and experiences. This process led us to the final version of the framework, as detailed in Figure 2.

---

[a] https://medlineplus.gov/
[b] https://www.cdc.gov/index.htm
[c] https://www.who.int/

*Co-design Workshops*

**Figure 1.** An illustration of our design process, which highlights the various stages involved in our design process.

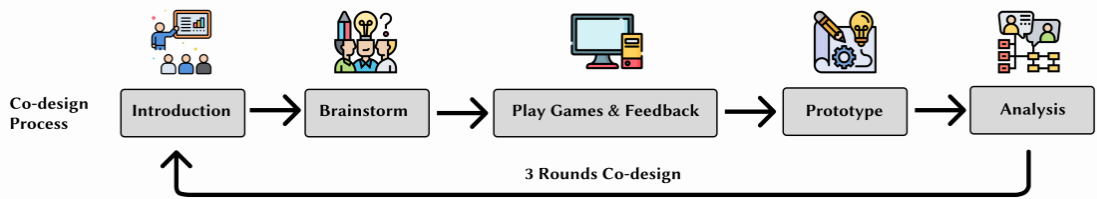

To address user needs in real-world contexts and refine the design space, we employed a participatory design approach, inviting individuals with diverse online meeting experiences to contribute to the development of gamified bodily interactions. We organized three online co-design workshops to elicit participants' meeting experiences, create new design ideas and interactive demos, and gather design insights. Eleven participants (3 males, 8 females) were recruited with diverse educational and professional backgrounds, including business, design, and engineering. Among them, seven were aged 18-24, two were aged 25-34, and two were aged 45-54. We refer to these participants as P1 - P11. Each workshop session lasted approximately two hours. To facilitate the co-design process, we prepared prompt cards to inspire participants during their discussions and idea sketches. These cards are designed based on our literature study and firsthand experience accumulated earlier.  They were presented at 8 card sets, reflecting our initial design considerations as mentioned above. Each set addressed a key consideration, including time to start, duration, body movements as input, bodily interplay, action, game element, attention, and exercise intensity. For example, the card set named "Action Cards" include 21 illustrations with each depicting a beneficial bodily exercise found in literature and credible online sources as aforementioned. As another example, the "Attention Cards" set includes a spectrum, ranging from games that allow you to play while focusing on the meeting to games that require you to fully concentrate on.

The workshops were hosted on Zoom, aiming to resemble an online meeting setting and allow the participants to test interactive demos within the Zoom interface. Design activities were completed using Figma and the FigJam feature. The co-design workshop was divided into five steps, as Figure 1 illustrates: introduction, brainstorming, gaming and feedback, prototyping, and iterating. Participants were introduced to the research background, purpose, and workshop schedule during the introduction stage. During brainstorming, participants shared their ideas, provided suggestions, and generated new concepts. The gaming and feedback stage involved playing gamification demos or acting out design ideas together and exchanging opinions in a group discussion. Inspired by the first three stages, participants sketched out their ideas during the prototyping stage. Finally, participants refined their prototypes with sharing their further considerations. After the workshops, we acquired a rich understanding of underlying real-world needs from the participants.

We collected two types of data during this process. Firstly, we amassed a significant number of ideas from the brainstorming phase of the co-design workshops. Secondly, we documented group discussion records. To further clarify the insights from idea cards, we employed an affinity diagram, helping identify common themes, patterns, and relationships among the data points. For discussion content, we utilized Hsieh's conventional qualitative content analysis method [57] to identify how initial design considerations supported participants in formulating their ideas.

**Insights from co-design workshops**: The co-design workshops yielded rich insights. We identified four major themes in the affinity diagram (see Table 1). First, the social aspect was identified as critical, with competition and cooperation significantly influencing user engagement. Second, participants expressed a preference for games with low learning curves. This is aligned with previous work by Mandryk et al [42] which proposed a design principle: *"providing an easy entry into play"*. The third point addressed the physical fatigue and mental stress present in online meetings, highlighting the need to differentiate between the intensity of physical activity and the attentional demands of games. Finally, we observed variations in the layout view and display order of video tiles among different users, with tile sizes varying based on the number of meeting participants. Fluctuations in the number of attendees, speakers, and layout settings within the meeting platform can impact the meeting interfaces. These factors should be considered when designing games.

**Table 1.** User needs identified from workshops.

| User Needs | Representative Responses from Participants |
| --- | --- |
| Consideration of Social Factors | - "When I see the name of one of my best friends on it, my desire to win is very strong." (P8)<br>- "I am more inclined towards something that can be played alone." (P6) |
| Desire for Simplicity and Familiarity | - During brainstorming, many participants suggested game ideas inspired by Fruit Ninja and Whack-a-Mole, explaining that "everyone is already familiar with the rules of these games." (P3) |
| Adaptability to Physical and Mental States | - "The mind is already very tired (during the meeting)." (P11)<br>- "If (the intensity of physical activity is) too gentle, I can't feel my body being relaxed." (P2) |
| Clear Visibility and Usability | - "I can't clearly see the sticker elements on the video." (P9) |

For the discussion content analysis, participants highlighted the significant utility of the design consideration cards we supplied in shaping and refining their ideas. They provided the following insights on how to use the cards: inspiring and scaffolding non-expert users to propose a design, adapting the design into the online meeting context, and evaluating the design. These findings also offer insights into the application of the proposed BIG-AOME framework.

*BIG-AOME Framework*

**Figure 2.** BIG-AOME (Bodily Interaction Gamification towards Anti-sedentary Online Meeting Environments) framework.

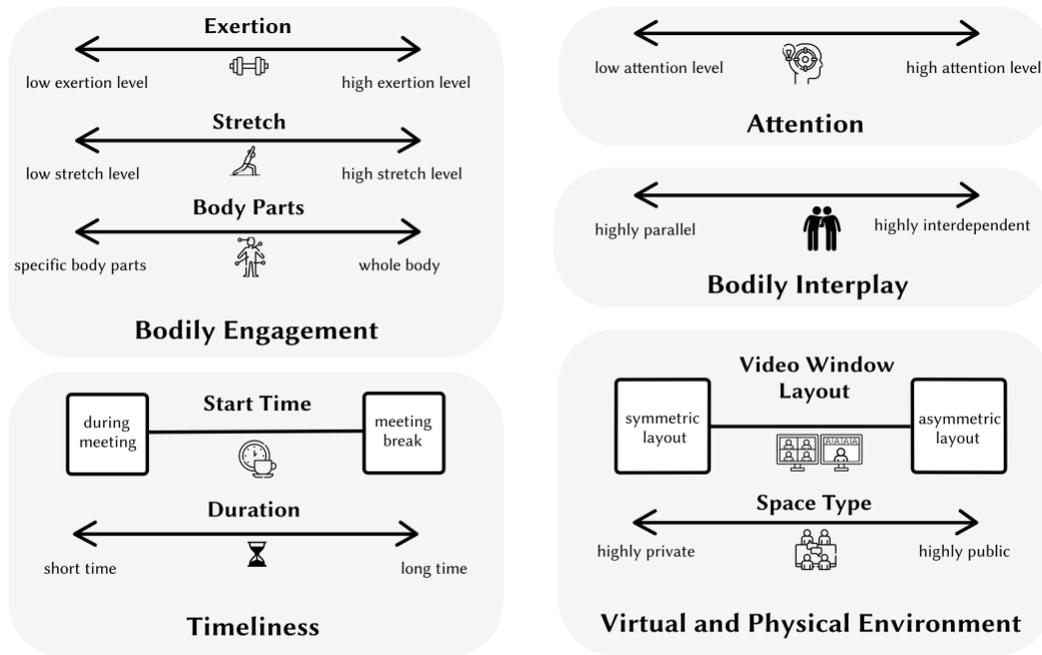

In this section, we present the BIG-AOME (Bodily Interaction Gamification towards Anti-sedentary Online Meeting Environments) framework. Initially formulated during early design explorations, the framework has been enriched and refined based on insights and design implications derived from the co-design sessions. As illustrated in Figure 2, the framework consists of five dimensions that specify a broad design space for creating gamified anti-sedentary bodily interactions for online meeting scenarios. Here we briefly introduce each dimension whereas a more concrete examination will be provided later using the empirical data gathered from the user evaluation.

*Bodily Engagement*: The *bodily engagement* dimension describes players' physical experience during gameplay. It includes three sub-dimensions: *exertion*, *stretch*, and *body parts*. *Exertion* represents the perceived intensity of physical effort users put forth during the game. *Stretch* refers to the perceived extent of physical reach and flexibility a game prompt. The *body parts* sub-dimension indicates the perceived extent to which various parts of the body are engaged during gameplay. This could range from games that primarily involve a specific part such as head-nodding, to more full-bodied experiences like dancing. Incorporating the three sub-dimensions allows us to design games that are more pertinent to the nuanced needs and the design goals.

*Attention*: The *attention* dimension measures the concentration needed for gameplay and highlights positive multitasking during online meetings. Games requiring peripheral attention enable users to move without losing focus on the meeting. Conversely, games requiring more focus provide high engagement and offer attendees a pause, promoting physical activity and fostering interactive play amongst participants. With numerous design nuances, the attention dimension offers a wide spectrum for game design, ranging from lightly engaging to fully immersive experiences.

*Bodily Interplay*: The *bodily interplay* dimension builds upon the concept proposed by Muller et al [41]. It evaluates to what extent users' bodies interact and influence each other during the gaming experience. High interdependent bodily interplay requires collective participation, fostering a shared gaming experience. In contrast, parallel bodily interplay allows participants to engage with the game independently, without requiring cooperation from others. This form of interaction is non-interfering and can be asynchronous. The level of bodily interplay is a vital design consideration, depending on the design objectives and the social dynamics of the meeting attendees. Therefore, it requires thoughtful deliberation in the design process.

*Timeliness*: The *timeliness* dimension comprises two sub-dimensions: *start time* and *duration*. The *start time* sub-dimension determines the most suitable moment to commence the game. Some games are designed to be engaged with while the meeting is ongoing, whereas others might be better suited to ad-hoc breaks. The *duration* sub-dimension focuses on the length of an episode of the gaming experience. Episodes can be brief, offering a quick gamified experience that ends once completed, or continuous, allowing intermittent engagement over a longer period. The design choices for both sub-dimensions depend on the meeting context and attendees' preferences, requiring careful consideration.

*Virtual and Physical Environment*: The dimension addresses two crucial factors: the layout of the video conferencing window and the nature of the user's physical space. Video conferencing software typically provides either symmetric (e.g., thumbnail video tiles) or asymmetric layouts (e.g., speaker view). In an asymmetric layout, typically utilized when a participant shares their screen, the shared content consumes a larger part of the display. This setup enables the shared screen to display gameplay elements to all participants, collectively drawing their attention to the game. The larger display area in this layout may afford more visual elements. Conversely, in a symmetric layout, the screen is equally divided. This layout enables the display of visual elements on each player's individual view, allowing them to see more faces. However, this layout may result in each user's video window becoming too small to clearly view smaller visual elements, particularly in large meetings. Designers should recognize the differences between symmetric and asymmetric layouts and consider the nature of the meeting. Turning to the physical environment, the *space type* pertains to the level of privacy and ownership of a user's surroundings. A highly private space, such as a personal study, enables users to move and speak freely without social concerns. Conversely, a highly public space, like a library or office, may restrict a user's behavior to avoid disturbing others or creating social awkwardness. These constraints can affect users' willingness and ability to engage in games during meetings.

### Three Design Prototypes

To explore how gamified bodily interactions can be integrated into online meetings to reduce sedentary behavior (RQ1) and to contextually examine and refine the BIG-AOME framework for relevant design implications (RQ2), we developed three high-fidelity prototypes for user evaluation: "Virus Hitter," "Frost," and "Food Rain," as shown in Figure 3.

**Figure 3.** Three high-fidelity prototypes implemented to probe the design of anti-sedentary bodily gamification for online meetings.

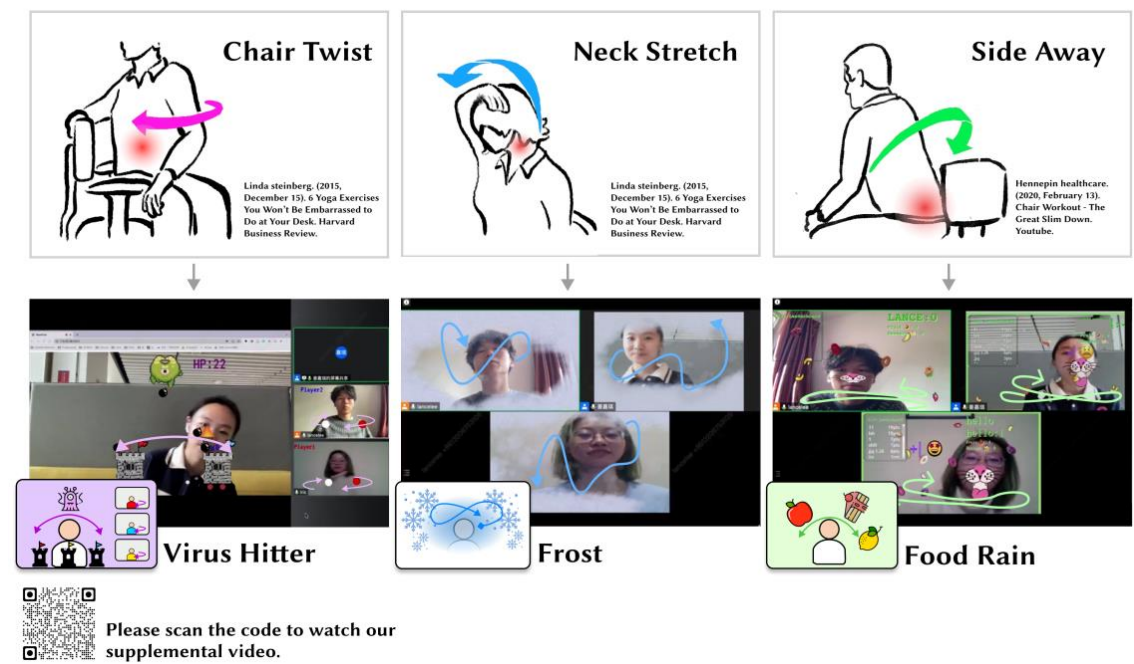

"Virus Hitter" is a multi-player cooperative game. In this game, players are assigned roles wherein one is randomly selected as the "hitter", while the rest serve as the hitter's assistants. The assistants' interaction is adapted from an existing exercise called *chair twist* [58]: participants while seated, are asked to rotate their upper bodies from one side to the other and repeat, facilitating a relaxing effect. The hitter then moved the torch attached to his/her nose to launch the bombs on the watchtowers. The hitter's interaction is derived from an exercise called *side sway* [59], which involves side-to-side sway while remaining seated in a chair. Every assistant corresponds to one watchtower matched by color. A brief animated guidance is offered so that players could quickly and easily understand how to play the game. "Virus Hitter" is designed to promote a high level of bodily interactivity and requires the cooperation of all participants, thereby not only encouraging physical activity but also fostering social bonding among players.

"Frost" is a gamified interaction designed for use during online meetings. It simulates frost slowly forming on the surface of the video window's glass, gradually spreading from the edges toward the user's image. To prevent their personal window from becoming entirely obscured by frost, users must move any part of their body to "swipe" it away. The suggested movement for this activity is inspired by the *neck stretch* [58], involving the simple action of raising and lowering the head to stretch the neck muscles. This specific interaction is chosen to provide a natural and inconspicuous reason for meeting participants to engage in light physical activity without feeling self-conscious. This game is characterized by low levels of exertion, stretch, and attention, making it ideal for relieving muscle and spine fatigue through gentle and unobtrusive movements. These movements do not significantly distract from meeting discussions or disrupt the flow of the meeting, whether in virtual or physical settings. It is designed to be performed seamlessly as a secondary task, complementing the primary activities of an online meeting, thus enabling participants to stay active without interrupting the meeting dynamics.

"Food Rain" is a multi-player competitive game. After inputting a nickname, fruits and desserts will fall off from the top of the screen. Players move their bodies and open their mouths to catch the falling foods in their own window to score. Catching fruits leads to an increase in the player's score, while catching desserts results in a reduction of the score. This movement is also derived from *side sway* [59]. During this process, players' neck and waist are exercised in moderate strength. Catching a fruit would earn one point while eating a dessert would minus one point. As opening the mouth in front of others might feel awkward to some users, the design superimposes a cartoonish animal mouth onto the user's mouth and can track the user's mouth to open and close accordingly. At the top left of user's window, there is a leaderboard displaying the ranking and score of all players. The competition mechanism encourages long-term participants. With middle level of duration, the game was designed with the assumption that it would trigger relatively high social engagement level.

The three gamified bodily interactions are implemented on a web-based application, streamed via Open Broadcaster Software (OBS) [55] as a virtual camera, which enables its integration within any online meeting platforms (e.g., Zoom and VooV Meeting). As the online meeting software and its application stores continue to evolve, gamified bodily interactions have the potential to emerge in various forms in the future. They could be available as add-on applications within the platforms, integrated as features of the meeting interface itself, or even implemented as virtual backdrops for meetings.

**Figure 4.** Distribution of dimensions on the framework when designing the three prototypes.

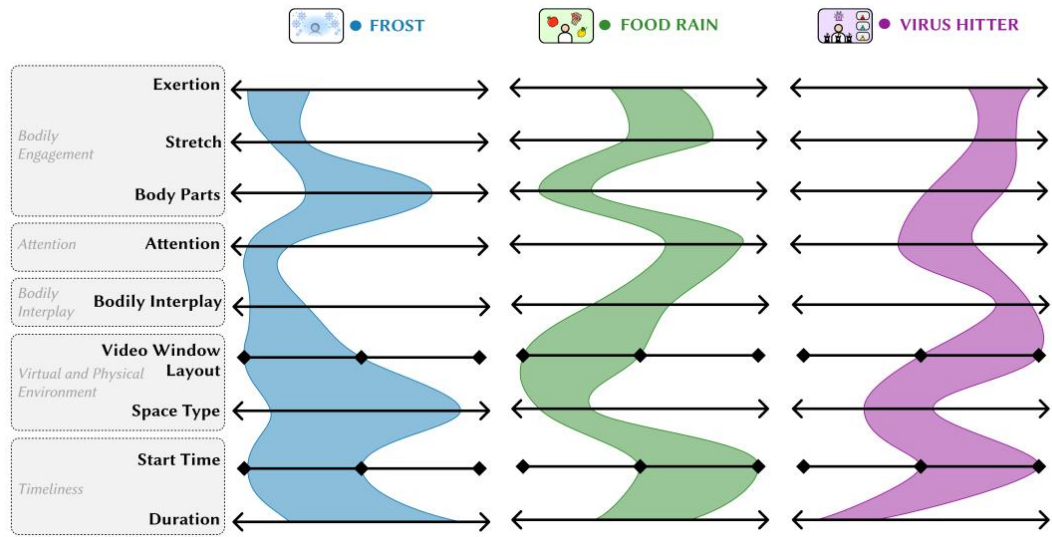

The prototypes presented here are representative examples chosen to illustrate and explore the design space. They were selected from a range of game designs developed during our early exploration and co-design phases. In the process of transforming ideas into prototypes, we considered various dimensions of the framework. As depicted in Figure 4, each prototype is intentionally situated in different regions within these dimensions. This strategic placement results in each prototype embodying unique design patterns. Each dimension contributes to the overall game design, and different combinations of dimensions can yield a diverse range of games.

## Evaluation Study

### Participants and Setup

The study involved three rounds of sessions with 4, 5, and 6 participants, totaling 15 participants (10 males, 5 females), referred to as P12 to P26. Among them, eleven were aged 18-24 and four were aged 25-34. A pre-study survey was conducted to gather participants' basic information and online meeting habits. The results showed that 53.3% of participants attended one or more online meetings weekly, and 86.7% spent more than an hour in each meeting. To simulate a realistic online meeting scenario, we organized three rounds of online seminars focused on ChatGPT. Each session included interactive discussions and integrated gamified activities. The seminars were designed as collaborative learning experiences, where participants were introduced to foundational ChatGPT concepts through lecture videos and group discussions on open-ended questions. Fixed-interval breaks were incorporated into the meetings, during which participants engaged with gamified exercises. After the seminar, participants independently evaluated the games by completing an evaluation panel designed using Figma. This panel, based on the BIG-AOME framework, aimed to capture their feedback on the gaming experience and design elements (see Multimedia Appendix 1 for details). Each session concluded with a focus group interview, where participants discussed their gaming experiences and shared insights into their evaluation choices (see Multimedia Appendix 2 for details). To ensure a smooth process, all games were pre-configured on participants' browsers, and clear guidelines were provided for the evaluation and interview stages. The entire meeting process, including the focus group discussion, was recorded for subsequent analysis.

**Figure 5.** Screenshots showcasing the user evaluation process (left) and the interview process (right) conducted on the online meeting platform.

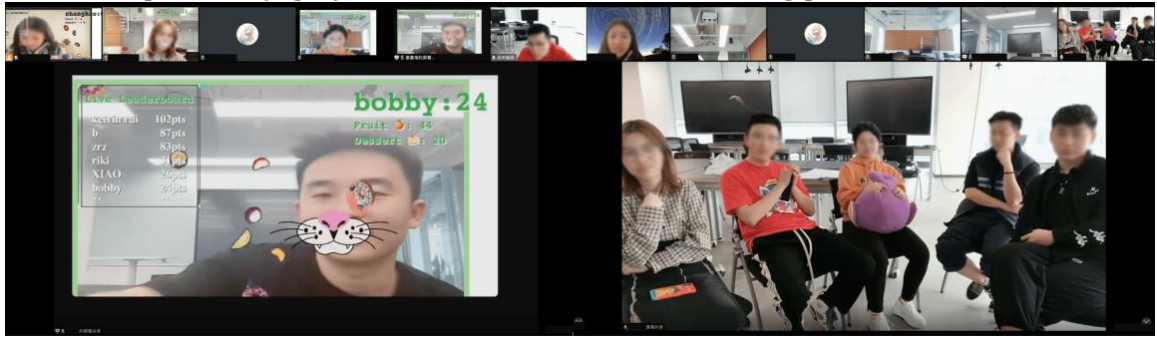

*Procedure*

As illustrated in Figure 6, the entire study spanned 110 minutes, comprising 45 minutes for learning activities, 30 minutes for gameplay, 10 minutes for completing the evaluation panel, and 20 minutes for a focus group interview. Each game was allocated 10 minutes, providing participants ample time to engage as they wished. The study began with the host introducing the study's objectives and process. Participants then watched a YouTube video explaining ChatGPT and its underlying mechanism. Following this, the first meeting break allowed participants to play one randomly selected game. After the break, participants were divided into groups to discuss the risk and opportunities brought by ChatGPT. Each group was assigned one team member to facilitate the discussion. Another game was randomly selected for all participants to experience. After playing, one representative in every group presented their opinions to others. The final game was played during the last meeting break.

After finishing the simulation of seminar, participants were invited to evaluate the games and propose their personal insights about how to apply the games in their real-life scenarios. Participants were given 10 minutes to finish the evaluation panel. We emphasized that participants should finish the evaluation panel based on their own experience and any result is acceptable. In the end, a 20-minute focus group interview was conducted to explore participants' experiences with gamified bodily interactions and discuss potential application scenarios. The group discussions were structured as semi-structured, employing open-ended questions to facilitate open and comprehensive conversations. First, we asked participants to recall their gaming experience, including physical sensations, mental responses, and interactions with others. Next, participants explained the reasons behind their markings on the evaluation panel. Finally, we invited participants to describe how they would utilize the framework to create a similar interactive game if given the opportunity.

**Figure 6**: A visualization of overall evaluation procedure.

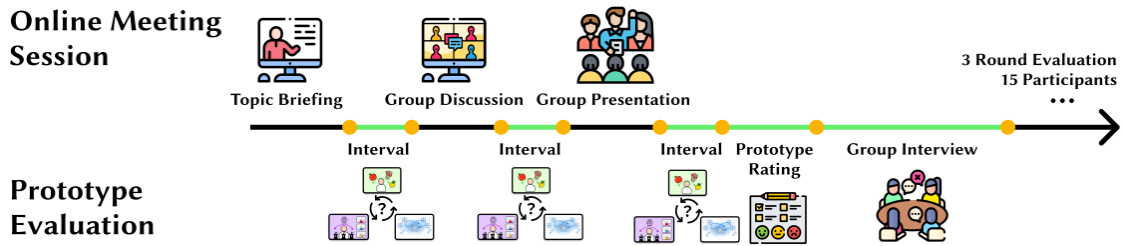

*Data Gathering and Analysis*
Data collection and analysis comprised two parts. First, we conducted focus group interviews to gather users' experiences with gamified bodily interactions during online meetings. All interview data were transcribed verbatim. We employed Hsieh's conventional qualitative content analysis method[57] to deeply understand participants' perceptions and experiences with the games used during meeting breaks. This process led to the development of preliminary codes, which were iteratively refined and adjusted as our understanding of the data evolved. These codes were then organized into clusters, providing a clear outline of the major findings from our qualitative data.  Second, participants rated each of the three games based on their experiences within the dimensions of the BIG-AOME framework. These ratings were instrumental in identifying the framework regions occupied by each prototype and in validating our initial design assumptions. The ratings were processed using Python and are descriptively visualized in Figure 7.

**Results**

**Integrating Gamified Bodily Interactions into Online Meetings (RQ1)**
The first aim of our study is to explore how gamified bodily interactions can be effectively integrated into online meeting settings (RQ1). This investigation seeks to provide empirical insights into this novel anti-sedentary approach, identifying potential opportunities and addressing challenges to inform future applications. Our findings include three aspects (see Table 2): (1) Perceived value to break sedentary behavior patterns; (2) Potential factors to motivate and sustain usage; (3) Anticipated usage scenarios and additional benefits.

**Table 2.** User needs identified from workshops.

| Key Themes | Detailed Description |
|---|---|
| Perceived value to break sedentary behavior patterns | - Reason to move.<br>- Reduces the awkwardness of moving during meetings. |

| Potential factors to motivate and sustain usage | - Seamless integration with meeting software.<br>- Long-term rewarding elements. |
|---|---|
| Anticipated usage scenarios and additional benefits | - Online ice-breaking activities.<br>- Gamified decision-making. |

*Perceived Value to Break Sedentary Behavior Patterns — "now I have a reason to move."*

Participants affirmed that the designed bodily interactions provided a compelling reason to move during online meetings, offering innovative ways to disrupt sedentary routines. Based on their prior experiences, they could all relate to the physical strain and *"discomfort"*（P21） caused by the sedentary nature of online meetings. Introducing bodily interactions, as P22 noted, could effectively reduce sedentary behavior, and encourage physical movement in online meeting scenarios. P22 expressed, *"(It brings) more physical activity in online meetings."* To some participants, when they were in an online meeting and wanted to move but felt *"embarrassed"*, *"This game gives me a chance, now I have a reason to move."* P23 described a similar moment in which he/she thought *"You often remain static during online meetings, but adding such games will add more subjective movement."* Also, P12 and P14 indicated that after experiencing the games they felt more relaxed: *"A little more relaxed than at the beginning for sure."* Overall, participants expressed that gamified bodily interactions not only gave them a reason to engage in physical activity but also reduced the awkwardness associated with moving during meetings.

*Potential Factors to Motivate and Sustain Usage —"This will motivate me to stick with it."*

Participants agreed that the games were effective in encouraging movement because they were novel, easy to learn, and facilitated highly interactive sessions that reduced social awkwardness. However, they acknowledged that the novelty and social benefits might diminish over time, particularly in routine meetings where attendees are already familiar with the games. To address this, participants emphasized the need for strategies to sustain engagement over the long term. This supports our view that gamified bodily interactions should be integrated into meeting routines rather than being transient or overly intense, promoting lasting behavioral change. Interviews revealed two key factors influencing user experience that are critical for achieving sustained success.

*Seamless integration with meeting software*: P19 suggested, *"I think a pop-up prompt might remind me to use it more."* Accordingly, it might be beneficial to incorporate reminder prompts on meeting platforms during breaks or between long sessions. However, the timing and delivery of these reminders need to be carefully considered. There is existing research indicating that notifications delivered at inappropriate times can distract users and potentially lead to annoyance [19]. P21 expressed that seeing others engaging in anti-sedentary interactions could motivate them to participate as well. With seamless integration, users could be effectively prompted to initiate the bodily interactions.

*Long-term rewarding elements*: Extrinsic motivation arises from external factors, such as rankings and individual records [60]. Implementing these motivators, such as points, badges, and leaderboards, in gamified bodily interactions can foster a competitive environment, thereby encouraging participants to remain engaged and enhance their performance. For example, P13 noted that during the game "Food Rain," competitiveness was heightened due to the presence of a leaderboard: *"We played pretty aggressively because of the leaderboard."* Additionally, P16 described the influence as akin to "peer pressure". Furthermore, offering a variety of games can cater to diverse participant preferences and prevent monotony. Regular updates to existing games, including introducing *"new game patterns"* (P19) and *"new visual and aural effects"* (P13), are both convenient and effective strategies for sustaining player involvement. Indeed, this encapsulates the rationale behind our research use of gamification: to enhance extrinsic motivations for physical activity and reduce sedentary behavior.

### *Anticipated Usage Scenarios and Additional Benefits ──"it could be used as an icebreaker activity."*

When discussing the potential application cases of gamified bodily interactions in online meetings, participants highlighted the possibility of integrating these games into various scenarios. Gamified bodily interactions can serve as icebreakers at the beginning or breaks of online meetings, helping participants to become acquainted with each other and create a comfortable atmosphere. These games were experienced to help break down communication barriers, especially in situations where participants had not met before or had limited prior interactions. For instance, P13 appreciated the game "Virus Hitter" as a potential means for group ice breaking. P17 and P24 also highlighted the potential of such interactive games serving as effective icebreakers in virtual settings *"it could be used as an icebreaker activity."* (P24) This insight underscores the versatile applicability of gamified bodily interactions in online meetings, extending beyond mere physical engagement to fostering social connections.

In online meetings where decision-making is required, games can be used to facilitate discussions, promote active participation, and encourage team members to share their opinions. The interactive nature of these games can help maintain engagement during lengthy discussions and ensure that all voices are heard. For instance, P13 envisions the integration of these games as a creative tool for participant selection while simultaneously energizing the meeting atmosphere: *"for example, to decide on a speaking order"*. In situations where a group needs to choose a member to speak or present but is unsure about whom to select, they can initiate the game and obtain a ranked list of candidates. Motivated by the ranking system and the prospect of being chosen to speak, participants are likely to engage actively in the game and incorporate physical activity into their meeting experience. This innovative approach to participant selection not only encourages physical movement but also adds an element of excitement and competition, fostering a livelier and engaging online meeting environment.

## Contextualizing BIG-AOME Framework (RQ2)

In this section, we contextually examine and concretize the framework with the empirical data gathered from evaluation sessions. As illustrated in Figure 7, participants were asked to position the three evaluated prototypes on each sub-dimension to verify initial design assumptions and gain an overview of how these design instances are distributed over this design space. Descriptive statistics are presented in Table 3. Using the gathered data, we analyze each design dimension and elaborate on relevant design choices, to surface the design space and considerations for future research.

**Table 3.** Descriptive statistics of user rating results across games and dimensions.

| Dimension | Game | Mean (SD) | Q1 [a] | Q3 [b] |
|---|---|---|---|---|
| Exertion | Food Rain | 0.462 (0.251) | 0.306 | 0.528 |
| | Virus Hitter | 0.551 (0.156) | 0.466 | 0.649 |
| | Frost | 0.207 (0.109) | 0.158 | 0.263 |
| Stretch | Food Rain | 0.474 (0.276) | 0.278 | 0.636 |
| | Virus Hitter | 0.609 (0.210) | 0.453 | 0.750 |
| | Frost | 0.349 (0.144) | 0.250 | 0.418 |
| Body Parts | Food Rain | 0.278 (0.172) | 0.157 | 0.430 |
| | Virus Hitter | 0.570 (0.212) | 0.419 | 0.750 |
| | Frost | 0.206 (0.220) | 0.090 | 0.236 |
| Attention | Food Rain | 0.781 (0.256) | 0.703 | 0.984 |
| | Virus Hitter | 0.523 (0.276) | 0.422 | 0.690 |
| | Frost | 0.228 (0.160) | 0.141 | 0.306 |
| Bodily Interplay | Food Rain | 0.414 (0.270) | 0.220 | 0.594 |
| | Virus Hitter | 0.858 (0.176) | 0.799 | 1.000 |
| | Frost | 0.178 (0.273) | 0.000 | 0.248 |
| Duration | Food Rain | 0.478 (0.271) | 0.337 | 0.601 |
| | Virus Hitter | 0.491 (0.167) | 0.390 | 0.573 |
| | Frost | 0.503 (0.270) | 0.282 | 0.669 |
| Space Type | Food Rain | 0.322 (0.310) | 0.08 | 0.395 |
| | Virus Hitter | 0.570 (0.311) | 0.330 | 0.784 |
| | Frost | 0.458 (0.340) | 0.225 | 0.722 |

[a] Q1 represents the first quartile (25% of the data falls below this value).
[b] Q3 represents the third quartile (75% of the data falls below this value).

### Bodily Engagement —How does the game involve bodies and challenge physically?

The first dimension of the framework focuses on the physical involvement and challenges presented by the game. This dimension encompasses three sub-dimensions: *exertion*, *stretch*, and *body parts*.

As shown in Figure 7, we can observe that the three prototypes follow a similar pattern in the three sub-dimensions: "Virus Hitter" > "Food Rain" > "Frost". This is in line with our initial design intentions. However, it's important to note that the ordering of games in the three sub-dimensions is not always consistent. For instance, a game involving multiple body parts can still require low exertion, such as matching body key points to a static shape — an idea proposed by the co-design participants.

**Figure 7**: Results of users' evaluation regarding the distribution of the three games on the framework. [a]

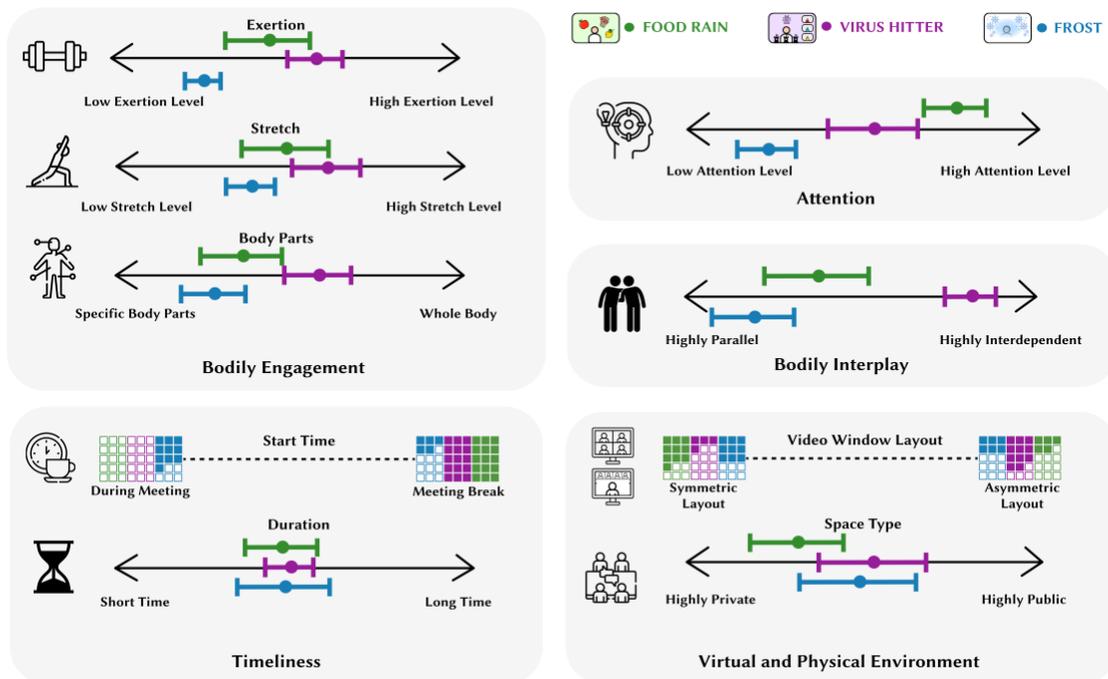

[a]The horizontal line for each game represents the interquartile range (Q1-Q3), defining the middle 50% of the data, which highlights the central distribution and concentration of the data values.

Designs with lower bodily engagement levels appeal to a wider range of participants and reduce fatigue, enabling players to engage more frequently or for longer periods. However, these designs may not effectively engage individuals who prefer high-energy activities or competitive environments, potentially leaving them feeling unchallenged or uninterested. In contrast, designs with higher bodily engagement offer more intense physical challenges. However, high bodily engagement games may not be suitable for all participants due to physical limitations or personal preferences, leading some individuals to feel excluded or frustrated if they cannot fully engage in the activities. For example, P19 expressed that *"I'm too tired"* when he/she experienced the game "Virus Hitter". P12 said, *"The last game ("Virus Hitter") ('s bodily engagement) is more obvious, the first two ("Frost" and "Food Rain") are relatively easy."* This highlights the varying levels of bodily engagement across the three games, confirming the noticeable differences in their physical challenges.

*Attention —What level of focus is required?*
It can be observed from Figure 7 that in terms of attention, "Food Rain" demands the highest level of attention, followed by "Virus Hitter", and then "Frost". The three games span almost the entire attention dimension, providing players with varying levels of cognitive engagement. This diverse range of attention levels can cater to the different preferences and scenarios in online meetings.

Bodily interactions that require low attention allow participants to engage without disrupting the meeting. They can maintain focus on the meeting content or personal tasks simultaneously, ensuring that the overall flow of the meeting remains uninterrupted while participating in peripheral bodily interactions. Additionally, incorporating a design that only requires peripheral attention during a break in the meeting can be an effective way to encourage relaxation and mental rejuvenation. Participants can engage in the game while also taking care of other tasks, such as drinking water. For example, P19 said that *"I can do something else (during this time of experiencing the gamified interaction)."* However, due to the peripheral nature, these games might not fully engage participants or provide a deeply immersive experience. P21 felt that "Frost" is *"not like a game"* and therefore has less enthusiasm to play with it. Games that require high attention can create a more engaging and immersive experience. Moreover, such games can offer a complete shift from the meeting content, allowing participants to refresh their minds before returning to the meeting activities. However, high attention level can lead to greater mental stress or cognitive load for participants.

*Bodily Interplay —To what extent can bodies act upon and react to each other in gamified bodily interactions in online meetings?*
As demonstrated in Figure 7, "Virus Hitter" demonstrates a significantly higher level of bodily engagement compared to "Food Rain" and "Frost". Collectively, the prototypes cover a wide range in this dimension.

Games with high parallelism provide flexibility in terms of participants' availability, allowing individuals to engage or disengage without impacting others. This feature is especially suitable for online meeting scenarios where participants may already be under mental strain. With minimal reliance on other players, individuals may feel less pressured to excel, fostering a more relaxed, enjoyable experience for those seeking less competitive interactions. However, such games may become repetitive and less captivating over time due to the limited participant interaction and challenge levels. Conversely, games with high interdependence can foster collaboration and communication, potentially strengthening team dynamics. Such games, particularly when incorporating competitive or cooperative elements, can enhance participants' engagement and motivation. Nonetheless, they also present challenges, as the players' enjoyment and success heavily depend on their teammates' or opponents' actions, which can impact the overall experience.

*Timeliness —When should the interaction begin, and how long should it last?*
As illustrated in Figure 7, regarding the preferred start time for each game, all participants suggested that both "Food Rain" and "Virus Hitter" are better suited to initiate during session breaks. Most participants (10/15) felt that "Frost" is more appropriate to initiate in the middle of the ongoing meeting activities. These insights correspond well with our design expectations, suggesting that our strategy to decide optimal game timings aligns with participant preferences.

Incorporating a game during a meeting break presents several benefits, including minimizing meeting disruption, facilitating social interaction among participants, and potentially enhancing team relationships. However, there are also drawbacks. Participants might prefer to use breaks for other tasks, such as attending to personal needs or engaging in work-related discussions. Furthermore, if some participants do not wish to participate in games during breaks, they might feel excluded or pressured to join. Conversely, initiating a game midway through a meeting offers a unique chance to invigorate participants. However, maintaining the right balance of attention is essential. Games incorporated into a meeting must be thoughtfully designed to avoid disrupting the flow or causing participants to lose focus on the main agenda.

Concerning the duration dimension, we noticed an intriguing pattern. All three games are positioned in the middle of the dimension, displaying minor differences. Our results indicate that user experiences across the three prototypes do not show considerable variations within the duration dimension. Short-duration games offer ease of integration into online meetings, serving as refreshing breaks that don't demand much time. They serve as effective transitions between meeting segments or as energizing preludes to more serious discussions. Offering frequent opportunities for physical and mental breaks, these games counteract sedentary behavior effectively. Long-duration games promise a more immersive experience with complex mechanics, deeper narratives, and enhanced team-building opportunities. However, fitting them into the meeting agenda without disrupting the flow can be challenging, and extended play might lead to participant fatigue if the game is too demanding.

*Virtual and physical environment —What are the videoconferencing layout and the level of physical space privacy?*
As shown in Figure 7, most people believe that "Virus Hitter" should be experienced in an asymmetrical layout, while "Food Rain" and "Frost" should be experienced in a symmetrical layout. This aligns with our design considerations.

An asymmetrical layout, where one participant shares their screen, provides an expanded display space, potentially enhancing the game's visual impact. However, this might inadvertently shift focus towards the sharer's content, potentially reducing interaction with other attendees or causing discomfort for the sharer under scrutiny. A symmetrical layout, with equally sized video windows for all participants, creates an equitable and harmonious environment, encouraging shared focus on self and others. However, the smaller video window size may impact game visibility and engagement, especially in larger meetings. Therefore, careful consideration and balance in the design process are essential to optimize user experience and participation.

Concerning the space type dimension, "Food Rain" is viewed as more suitable for private spaces, while "Virus Hitter" and "Frost" inhabit the middle ground. Playing in a private space offers comfort and security, allowing for uninhibited movement and gameplay without fear of social awkwardness or disturbing others. However, this may reduce social interaction and potentially lower motivation due to lack of group dynamics. In contrast, playing in a public space enhances social interaction and participation, fostering a sense of community and group camaraderie. The visible engagement of others can encourage individual participation. However, public spaces can lead to feelings of self-consciousness during gameplay, potentially affecting enjoyment and immersion.

## Discussion

### Design implications and opportunities towards anti-sedentary online meeting environments

Our study aligns with previous works by Warburton et al [61] and Aldenaini et al [15], which have demonstrated the benefits of gamification and persuasive technology for anti-sedentary designs in general. In the context of online teaching, Shin et al [45] and Sachan et al [46] conducted pilot studies to combined bodily interactions with online classroom scenarios. We further explored the vast design space and surfaced its underlying design implications. Furthermore, contributing to a design-oriented approach to this domain, we created prototypes for evaluation, enabling us to gather real-world experiences of participants within an online meeting environment. Here we discuss four key design implications that emerged from our study.

#### Providing proper and playful reasons for moving body during online meetings

Sedentary behavior is regulated by personal habit strength [62], and is often socially/environmentally reinforced [13]. Many new technologies are designed to reduce sedentary time, including games, reminders, wearable devices, and smart office equipment [42,63–65]. These interventions typically adopted action planning as the key mechanism. However, online meeting is a unique context where users are under additional stressors such as the closer-up "face-to-face" communication and reduced mobility [47], or distractions at home and online [24]. Participants also reported that they feel too embarrassed to stand up or move their body as they want even if they do feel very tired. In this situation, compared with a prompt of time for break, an incentive cue and a compelling reason are what meeting participants need more.

Medical evidence shows that the quantified threshold for sedentary time varies as health condition, age, gender, etc. [13]. There are no rigorous recommendations about the optimum limit to the sedentary time [66]. A study investigating the motivational processes underlying sedentary behaviors shows that action planning has a conditional effect on physical activity but no effect on limiting sedentary behavior [62]. This suggests that the focus should perhaps not be on setting strict activity goals or tracking exercises. Instead, encouraging movement and breaking up sedentary patterns whenever possible may be more beneficial. This does not necessitate high-intensity training or strenuous workouts. Even light or moderate movements incorporated into online meetings can already make a significant difference [67]. From a design research perspective, the challenge lies in creating socially engaging designs that provide meeting attendees with enjoyable, playful, and socially motivating reasons to move. Rather than strictly monitoring and regulating activity, the goal is to create an environment that naturally encourages movement and reduces sedentary behavior.

### *Designing unobtrusive bodily movement as secondary tasks to leverage positive multitasking in online meetings*

Multitasking is a common behavior in online meetings, and it can yield both positive and negative impacts [25]. Incorporating non-distracting, low-effort physical movements as secondary tasks during online meetings may serve as an effective design strategy for positive multitasking. Games with varying attention requirements offer diverse options for attention management, enabling participants to fluidly transition between relaxation and meeting focus. For instance, games that demand minimal attention can serve as a peripheral activity during the meeting. One such example is "Frost," which features a minimalistic visual interface and is intentionally non-distracting, making it suitable as a secondary task. The level of required attention is a crucial consideration in game design. By consciously defining the design goals and leveraging the benefits of positive multitasking, we can foster an anti-sedentary environment within online meeting routines.

### *Aiming for fine-grained integration with routines, rather than intensity or volume of physical activity*

Our work presents a unique proposition that deviates from conventional exertion game design aimed at sports training or physical activity tracking. In addressing sedentary behaviors during online meetings, our emphasis is not on the intensity or volume of physical activity. Instead, we prioritize designing interactions that are mild, moderate, easily initiated, and accessible to most participants. These interactions allow for flexibility, enabling users to engage with or dismiss them as needed.

In terms of mitigating sedentary behaviors, as prior research argues, the goal is not merely fulfilled by the increase of exertion intensity, but also the incorporation of light physical exercises [68] into the daily working scenarios. It is significant to realize that sedentary behavior is completely different from lack of physical activities. Most current exertion games are augmented sports [33] that rely on game consoles and sensor equipment to create a game-like sports experience, thereby promoting physical activity[36]. Our work represents a different approach to exertion games, designed specifically as playful anti-sedentary interventions. Instead of utilizing these games solely for workouts or ad-hoc entertainment, our objective is to seamlessly integrate them into existing online meeting routines. This integration aims to foster more lightweight, healthy behaviors among participants during their regular online engagements [42,69].

*Expanding design variety and integrating into existing meeting platforms to enhance real-world impacts*

In our exploration of the potential for gamified physical interactions within online meeting scenarios, we introduce the BIG-AOME framework. This framework serves as an initial foray into a five-dimensional design space, revealing a broad spectrum of design opportunities that can guide the creation of a diverse array of experiences tailored to various user preferences. Our objective with the BIG-AOME framework is to equip both researchers and designers with a foundational tool to further investigate this innovative direction. By promoting the integration of exertion games [33] with persuasive interventions [14], we aim to inspire a collaborative effort in the development of engaging and health-promoting solutions tailored for online environments.

Recognizing the diversity in user preferences and motivational factors, it is vital to adopt a personalized approach to game-based health interventions. An individual's engagement with the game is influenced by factors such as motivation, interests, athletic abilities, social connections, and social status [70]. Hence, to inspire long-term positive behavior changes, it is crucial to craft inclusive and engaging experiences that resonate with the varied expectations of diverse users. During the evaluation phase, we noticed a diversity of user preferences for the different games, which signals a degree of success in catering to our audience's varied expectations. Nevertheless, we acknowledge that numerous areas across the various dimensions and subdimensions remain untouched. This underscores the need for more diversified, design-driven research. By generating new design instances and extracting new design implications, designs in this domain could evolve towards more personalized, engaging, and effective game-based interventions.

Moreover, the importance of seamless integration with existing meeting platforms cannot be overstated. This echoes the principle of *"Providing an Easy Entry into Play"* proposed by Mandryk et al. [42]. Given the social nature of online meetings, users are often motivated to engage when they observe other users' usage on the platform. This affords new ways to increase user involvement [41]: with the growing ubiquity of online meetings, these platforms naturally present opportunities for an increased presence of gamified bodily interactions. By incorporating these interactions into existing meeting software and delivering appropriate signifiers about this new feature, users could become more conscious of their sedentary habits and are consequently more likely to partake in physical activities.

## Limitation and Future Work

While our study contributes valuable insights into gamified physical activities within online meetings, it is essential to acknowledge certain limitations that necessitate further research. Firstly, the evaluation period of our study was relatively brief, which may not adequately capture the long-term effects of these interventions. Future research should include longer evaluation periods to investigate the sustainability and long-term impact of such gamified interventions more thoroughly. Secondly, although we developed three high-fidelity prototypes that varied across different dimensions, these did not comprehensively explore some sub-dimensions. Consequently, while the data collected offers significant insights, these limitations may restrict the generalizability and robustness of our findings. Despite these constraints, we are optimistic about the extensive and promising potential of gamified bodily interactions in online meetings. Future studies could aim to experiment with more diverse and inclusive design features covering a wider range of physical activities. Expanding the scope of the interventions to include a broader variety of movements and postures will enhance our understanding of the potential and effectiveness of integrating gamified physical activities into online meeting environments. Additionally, investigating how gamified bodily interactions can stimulate users' intrinsic motivation to support sustained engagement over the long term is another valuable direction for future research.

## Conclusion

We utilized a research-through-design methodology to craft and explore the possibility of gamified bodily interactions as anti-sedentary interventions within online meeting contexts. In collaboration with 11 users, we co-designed and iterated three prototypes, which led to the development of BIG-AOME framework. Utilizing these prototypes, user studies were conducted with three groups totaling 15 participants. Empirical findings were gathered to understand user experiences with these prototypes and concretize the framework. Research findings indicate that designing anti-sedentary bodily interactions for online meetings has the potential to alter sedentary behaviors while enhancing social connections. Furthermore, the BIG-AOME framework proposed explores the design space for anti-sedentary physical interactions in the context of online meetings.